\documentstyle[epsf]{espart}

\begin{document}
\newfont{\riesig}{cmbx12 scaled 1400}
\newfont{\abklein}{cmr10 scaled 1000}
\newfont{\authors}{cmr12 scaled 1000}
\newfont{\adress}{cmti10 scaled 1000}

\def\klung{{\lower 2pt \hbox{$<$} \atop \raise 1pt \hbox{$ \sim$} }}
\def\grung{{\lower 2pt \hbox{$>$} \atop \raise 1pt \hbox{$ \sim$} }}

\begin{frontmatter}
\title{Semiclassical Density Matrix Near the Top \\ 
       of a Potential Barrier}
\author{Franz Josef Weiper},
\author{Joachim Ankerhold} and \author{Hermann Grabert}  

\adress{Fakult{\"a}t f{\"u}r Physik,
        Albert--Ludwigs--Universit{\"a}t Freiburg, \\ 
              Hermann--Herder--Stra{\ss}e 4, 
              79104 Freiburg, Germany}

\begin{abstract}
Employing the path integral approach, we calculate the 
semiclassical equilibrium density
matrix of a particle moving in a nonlinear potential field 
for coordinates near the top of a potential barrier. 
As the temperature is decreased, near a critical temperature $T_c$
the harmonic approximation for the 
fluctuation path integral fails.
This is due to a caustic arising at a bifurcation
point of the classical paths.
We provide a selfconsistent scheme to treat the large 
quantum fluctuations leading to a nonlinear fluctuation 
potential.
The procedure differs from methods used near caustics of the 
real time propagator.
The semiclassical density matrix is determined explicitly for 
the case of asymmetric barriers from high temperatures down to 
temperatures somewhat below $T_c$. 
\end{abstract}
\end{frontmatter}

\section{Introduction}
Processes hindered by potential barriers play a central role
in many areas of physics and chemistry  
\cite{haenggi1}.
When studying these systems, as a first step one has 
to determine the static behavior,
that is the equilibrium density matrix $\rho_\beta
\propto \exp(-\beta H)$.
In barrier penetration problems
the barrier height of the system is often
large compared to the quantum mechanical energy level spacing.
Then, a semiclassical evaluation of the equilibrium 
density matrix is justified.
A consistent scheme to determine the coordinate representation 
of $\rho_\beta$
in the semiclassical approximation is provided by the path 
integral approach \cite{p:feyn,p:schul,p:kleinert}.\\
Formally, the equilibrium density matrix 
may be viewed as a propagator in imaginary time.
Hence, one first has to investigate the classical 
mechanics in the inverted nonlinear potential $-V(x) $.
Afterwords, the quantum fluctuations about the classical paths
are determined.
Since the classical equation of motion, following
from Hamilton's principle of least action, has to be 
solved with fixed endpoints, one finds in general a 
set of classical paths contributing to the path integral.
If the classical paths, i.e.\ the minima of the classical 
action, are well separated in function space, the 
contribution of the quantum fluctuations around each path 
is given by the simple semiclassical approximation
where the fluctuation path integral is evaluated in the 
Gaussian approximation.
However, specific divergences, known as the problem 
of caustics, arise near those points 
where new classical paths become available.
For a system with a potential barrier one encounters a caustic
when the temperature is lowered, specifically in the temperature
range where thermally activated barrier crossing changes over to 
quantum tunneling \cite{gow}.
In this region one has to go beyond the Gaussian 
approximation for the fluctuation path integral.
For the case of a symmetric potential field, 
the semiclassical approximation near the caustic was 
investigated in a previous paper \cite{ankerhold}.
Here, we extend the results to asymmetric barriers.

The paper is organized as follows.
In section 2 we give a brief introduction to the path 
integral representation of the density matrix and the semiclassical 
approximation.
In section 3 we then determine the classical paths and the 
classical action of a particle moving in the inverted barrier
potential with endpoints in the vicinity of the barrier top.
In section 4 we study the quantum fluctuations and determine 
the semiclassical density matrix at high temperatures and 
for temperatures near the caustic.
Finally, in section 5 we give some explicit results for a barrier
potential with cubic and quartic anharmonicities and present our 
conclusions.

\section{Path Integral Representation of the Density Matrix
\hfill \break and Semiclassical Expansion}

The coordinate representation of the unnormalized equilibrium density 
matrix  of a particle of mass $m$ moving in a potential $V(x)$ may be 
written as \cite{p:feyn,p:schul,p:kleinert}
\begin{equation}
\rho_\beta(x,x^\prime) = \int \! {\cal D}[x] \; {\rm e}^{-\frac{1}{
\hbar} S_E[x]} ,  \label{eq:p19}
\end{equation}
where the functional integral is over all paths $x(\tau)$, $0 \leq 
\tau \leq \hbar \beta$ with $x(0) = x$, $x(\hbar \beta) = x^\prime$. 
Each path is weighted by its Euclidian action
\begin{equation}
S_E[x] = \int\limits_{0}^{\hbar \beta}\!\!  d\tau\, \left[ \frac{1}{2} 
m \dot{x}^2 + V(x) \right] .  \label{eq:p20}
\end{equation}
Since the density matrix can be regarded as the analytic continuation 
of the real time propagator to imaginary times $t \rightarrow -
i\hbar\beta$, the representation (\ref{eq:p19}) is frequently 
called the imaginary time path integral. The Euclidian action 
(\ref{eq:p20}) describes the real time motion of a particle in 
the inverted potential $-V(x)$.

Evaluation of the path integral (\ref{eq:p19}) in an asymptotic 
expansion for $\hbar \rightarrow 0$ provides a consistent scheme for 
a semiclassical approximation. There\-by one first determines the 
maximum of the weighting factor, that is the minimum of $S_E[x]$. 
This is given by the classical action $S_E[x_{\rm cl}]$, where 
$x_{\rm cl}$ is the classical path solving the classical equation 
of motion following from Hamilton's principle $\delta S_E[x]=0$. 
An arbitrary path in (\ref{eq:p19}) reads
\begin{equation}
x(\tau) = x_{\rm cl}(\tau)  + y(\tau)  ,\label{eq:p21}
\end{equation}
where $y(\tau)$ describes the quantum fluctuations about the 
classical path. The fluctuations have to fulfill the 
boundary conditions $y(0)=y(\hbar\beta)=0$. Using  
(\ref{eq:p21}) the full action is then expanded around its 
minimum according to
\begin{equation}
S_E[x] = S_E[x_{\rm cl}] + \sum_{n=2}^{\infty}\; \frac{1}{n!}\, 
\delta^n S_E[x_{\rm cl},y]   \label{eq:p22}
\end{equation}
with the Fr\^{e}chet derivatives
\begin{equation}
\delta^n S_E[x_{\rm cl},y] = \int\limits_{0}^{\hbar\beta}\! 
\!d\tau_1 \cdots d\tau_n  \left. \frac{\delta^n S_E[x]}{\delta 
x(\tau_1) \cdots  \delta x(\tau_n)}\right|_{x=x_{\rm cl}} \; 
y(\tau_1) \cdots y(\tau_n) . \label{eq:p23}
\end{equation}
This way the dominant term (for $\hbar \rightarrow 0$) is 
separated off and one is left with a functional integral over 
closed paths. In the simple semiclassical approximation, 
the sum  is truncated after the quadratic term leading to an exactly 
solvable Gaussian path integral \cite{p:feyn,p:schul,p:kleinert}. 
If there exists a set $\{x_{\rm cl}^\alpha \}$ of classical 
trajectories in $V(x)$, the above procedure must be performed for 
each $x_{\rm cl}^\alpha$, and all contributions are summed to 
yield the semiclassical density matrix
\begin{equation}
\rho_{\beta}(x,x^\prime) = \sum_{\alpha}\, 
\frac{1}{\sqrt{J_{\alpha}}}\, {\rm e}^{-\frac{1}{\hbar} 
S_E[x_{\rm cl}^\alpha]}  ,  \label{eq:p24}
\end{equation}
where $J_{\alpha}= {\rm det}\{\delta^2 S_E[x]/\delta x(\tau_1) 
\delta x(\tau_2)|_{x=x_{\rm cl}^\alpha}\}$ is the determinant 
describing the Gaussian integral over the quantum fluctuations 
\cite{p:schul,p:kleinert}.
Clearly, the result in (\ref{eq:p24}) is exact if $V(x)$ is 
harmonic. 
 $J_\alpha$ is given by the product of the eigenvalues 
$\Lambda_n^\alpha$ of the second order variational operator  
$\delta^2 S_E[x]/\delta x(\tau_1) \delta 
x(\tau_2)|_{x=x_{\rm cl}^\alpha}$ as
\begin{equation}
J_\alpha = N \frac{2 \pi \hbar^2 \beta}{m} \; 
\prod_{n}\,  \:\Lambda_n^\alpha ,
\label{eq:p24b}
\end{equation}
where $N$ is an appropriate normalization constant.      
As long as the second order variational operator  is 
positive definite, i.e. $\Lambda_n > 0$ for all $n$, the 
Gaussian approximation gives the leading order fluctuation 
term for $\hbar \rightarrow 0$.  But a problem arises if 
one of the eigenvalues $\Lambda_n$ tends to zero, e.g.\ as the 
temperature is lowered.  Then, the quantum fluctuations of this 
mode become arbitrarily large and the simple semiclassical 
approximation breaks down. Generally, the vanishing of an 
eigenvalue $\Lambda_n$ defines a point where new minimal action 
paths in the potential $V(x)$ become possible. This is  
well--known as the problem of caustics.
In those regions where a caustic arises one has to go beyond the 
simple semiclassical approximation and has to consider higher order 
terms in the expansion (\ref{eq:p22}) of the action.

\section{ Classical Mechanics near the Barrier Top }
To determine the density matrix in the semiclassical approximation 
we first have to study the classical mechanics of the system 
in the inverted potential.
Hence, we have to solve the classical equation of motion with the 
boundary 
conditions
$x(0)=x$, $ x(\hbar \beta)=x'$, where the endpoints $x$ and $x'$ 
are in 
the barrier region.
A general barrier potential may be written as 
\begin{equation}
V(x)= -\frac{1}{2}M\omega^2 x^2 \left[ 1- 2\sum_{n=3}^\infty 
\frac{a_n} 
{n} \left(x/x_a\right)^{n-2} \right] ,
\label{pot}
\end{equation}
where the $a_n $ are dimensionless coefficients that are of 
order 1 
or smaller and $x_a$ is the characteristic distance from the barrier 
top at which anharmonic terms of the potential become relevant. 
We assume that for coordinates in the vicinity of the barrier top
the potential (\ref{pot}) is only weakly affected by anharmonicities.
This means that the length scale $x_a$ is much larger
than the quantum mechanical length scale
\begin{equation}
x_0=\left(\frac{\hbar}{2M\omega}\right)^{1/2}
\label{x0}
\end{equation}
which is the variance of the ground state of an harmonic oscillator
in the inverted potential. Correspondingly, 
\begin{equation}
\epsilon = \frac{x_0}{x_a}
\label{delta}
\end{equation}
is a small dimensionless parameter which will be serve as
an expansion parameter for the semiclassical approximation
in the following.
This investigation becomes more transparent
if we introduce the dimensionless coordinate
\begin{eqnarray}
q &= & x/ x_0 ,
\end{eqnarray}
and the dimensionless time and inverse temperature
\begin{eqnarray}
\sigma &= & \omega \tau,\nonumber \\
\theta &= & \omega \hbar \beta .
\label{coord}
\end{eqnarray}
In terms of these dimensionless variables the scaled action 
$S=S_E/\hbar$ reads
\begin{equation}
S[q,q']= \frac{1}{4}\int\limits_0^\theta {\rm d} \sigma 
\left[\dot{q}^2(\sigma)-q^2(\sigma)\left( 1-2\sum _{n=3}^\infty \frac{a_n}{n} 
\epsilon^{n-2} 
q^{n-2}(\sigma)\right)\right].
\label{action}
\end{equation}

To determine the classical paths in the time interval $0 \leq 
\sigma \leq \theta $, it is convenient to use the Fourier series 
expansion
\begin{equation}
q(\sigma)=\frac{1}{\theta }\sum_{k=1}^\infty Q_k 
\sin (\nu_k \sigma)
\label{ansatz}
\end{equation}  
with the frequencies
\begin{equation}
\nu_k = \frac{\pi k}{\theta}.
\end{equation}
The above series continues the path outside $[0,\theta ]$ as an 
antisymmetric and periodic path with period $2 \theta $. The 
continued path has jump singularities at the endpoints of the 
interval $[0,\theta ]$.
This artefact of the Fourier series expansion must be taken into 
account when calculating time derivatives of $q$. 
Hence, in (\ref{action}) one has to insert $\dot{q}_{\rm reg}$
instead of $\dot{q}$ where
\begin{equation}
\dot{q}_{\rm reg}(\sigma) = \dot{q}(\sigma) - c_1 :\delta (\sigma, 2 \theta) : - c_2 : 
\delta (\sigma +\theta , 2 \theta):.
\label{regula}
\end{equation}
Here, the coefficients $c_1,c_2$ are determined by the jumps at $2 
k \theta $ and $(2 k-1)\theta $ as
\begin{eqnarray}
c_1 &= &q(0^+)-q(0^-)=2 q \nonumber \\
c_2 &= &q(\theta^+)- q(\theta^-)=-2q' 
\label{jumps}
\end{eqnarray}
and 
\begin{equation}
:\delta (\sigma, p) : = \sum _{k=-\infty}^\infty \delta(\sigma - k p)
\label{delta2}
\end{equation}
is a periodically continued $\delta $-function. 
Inserting (\ref{ansatz}) and (\ref{regula}) into (\ref{action}),
the action is found to read in terms of the Fourier amplitudes
\begin{eqnarray}
S(q,q') =& &\frac{1}{4 \theta}
 \Bigg[   (q-q')^2  \nonumber \\
& +&  {\frac{1}{2}\sum_{k=1}^{\infty }
\left(\lambda_k Q_k^2 -2 b_k(q,q') Q_k + 4\left[q-(-1)^k q'\right]^2 \right)}
   \nonumber   \\
& +&  { \sum_{n=3}^\infty \frac{a_n}{n} 
\left(\frac{\epsilon}{\theta}\right)^{n-2}
\sum_{k_1\dots k_n=1}^\infty D_{k_1 \dots k_n}Q_{k_1}\cdots Q_{k_n}}
\Bigg] ,
\label{action1}
\end{eqnarray}
where we have introduced the coefficients 
\begin{equation}
D_{k_1 k_2 \cdots  k_{n}} = 2\int\limits_0^1 
{\rm d}x
\prod_{j=1}^n \sin \left( \pi k_j x \right) ,
\label{d's}
\end{equation} 
and
\begin{equation}
\lambda_k =  \left( \frac{\pi k}{\theta} \right)^2 -1,
\label{eigenw1}
\end{equation}
as well as
\begin{equation}
b_k(q,q') =  \frac{2 \pi k}{\theta }\left[q- (-1)^k q'\right].
\label{eigenw2}
\end{equation}
Requiring that the variation
$\delta S[Q_k ]= 0$, we get the Fourier representation of 
the classical equation of motion
\begin{equation}
 \lambda_k Q_k = b_k - \sum_{n=3}^\infty a_n 
 \left( \frac{\epsilon}{\theta} \right)^{n-2}
 \sum_{k_1 ,\cdots ,k_{n-1} =1}^\infty 
 D_{k_1 k_2 \cdots  k_{n-1} k}
 Q_{k_1} \cdots Q_{k_{n-1}}.
\label{bewgl}
\end{equation}

In general, the nonlinear equations (\ref{bewgl}) cannot be 
solved exactly. 
However, for small $\epsilon$ and not too large $\theta$ one may solve 
them pertubatively in a selfconsistent manner.

\subsection{ High Temperatures} 
Since the boundary values $q,q'$ are in the barrier 
region and assumed to be of order 1 or smaller, 
for high temperatures, $\theta \ll 1$, the $b_k(q,q')$ are of order 
$1/\theta$ or smaller.
The anharmonic terms in (\ref{bewgl}) 
are then at most of order $\epsilon \theta $ and the classical equation 
of motion reduces to 
\begin{equation}
\lambda_k Q_k =b_k + {\cal O}\left( \epsilon \theta \right).
\label{harm}
\end{equation}
Neglecting the corrections, we obtain the Fourier representation 
of the minimal action 
paths of an harmonic oscillator.
When (\ref{harm}) is inserted into (\ref{action1}), 
the corresponding action is found to read
\begin{equation}
S(z,r)= 
       - \frac{r^2}{2}           \tan \left( \frac{\theta}{2} \right) 
       + \frac{z^2}{8} \cot \left( \frac{\theta}{2} \right)
       +  {\cal O} \left( \epsilon \theta^2 \right) ,
\label{harmact}
\end{equation}
where we have introduced dimensionless sum and difference coordinates
\begin{equation}
r=\frac{q+q'}{2},~z=q-q'.
\label{rx}
\end{equation}

This approximation is only valid for endpoints within the barrier 
region and temperatures where the amplitude of the 
classical path  remains within this domain.
Clearly, the harmonic approximation fails if one of the coefficients 
$\lambda_k$ vanishes.
When the temperature is lowered $\lambda_1$ reaches zero at the 
critical temperature
\begin{equation}
T_{c} = \frac{\hbar \omega}{ \pi k_{\rm B} },
\label{tc}
\end{equation}
i.e.\ $\theta_c=\pi$, and the corresponding amplitude $Q_1$ 
diverges.
Corresponding singularities of the harmonic approximation
arise for all temperatures where one 
of the frequencies $\nu_k=1$. This is not surprising, 
rather it reflects the typical behavior of a harmonic oscillator in 
real time.
For a harmonic potential the oscillation period $2\pi $ is 
independent of the amplitude of the path. Hence, for times 
$\theta =n\pi $, corresponding to multiples of half the oscillation 
period, there are infinitely many paths with arbitrary large 
amplitudes connecting $q$ and $q' =(-1)^n q$.

\subsection{Classical Paths Near $T_c$}

When the temperature approaches $T_c$ from above, the eigenvalue 
$\lambda_1$ tends to zero and the harmonic approximation
fails.
Then,  
the anharmonic terms in (\ref{bewgl}) become 
important. 
In general, the equation of motion cannot be solved 
analytically near $T_c$. 
However, the classical paths and their actions can always
be computed numerically. Here, we consider 
suitable potentials (see below) where only the amplitude $Q_1$ 
increases. 
For the $b_k(q,q')$ are now of order one or smaller and for 
perturbatively small $\epsilon $ we shall assume that for 
temperatures near $T_c $ the amplitude $Q_1$ may become at most 
of order $\epsilon^{-2/3}$, 
while all other amplitudes are of order 1 or smaller.
This order of magnitude of the Fourier coefficients will be confirmed 
below for appropriate barrier shape.
The equation of motion (\ref{bewgl}) then reduces to
\begin{equation}
\lambda_k Q_k = b_k -a_3 \frac{\epsilon }{\theta } D_{11k} Q_1^2
                    -a_4 \left(\frac{\epsilon }{\theta }\right)^2 
D_{111k} Q_1^3 +{\cal O}\left(\epsilon^{2/3}\right) .
\label{kub1}
\end{equation}
Within this approximation the amplitudes $Q_n$ for $n>1$ 
are coupled to $Q_1$ only.
We have to solve a cubic equation for the mode amplitude
$Q_1$ and  linear equations for the other mode amplitudes  $Q_n$.
To derive (\ref{kub1}) we have made two assumptions.
First, the parameter $a_4$ is assumed to be positive and of order 1.
In fact, the lenghtscale $x_a$ in (\ref{pot}) can always be chosen
such that $a_4=1$ unless $a_4$ is not positive.
Second, the coefficient $a_3$ can be taken positive without loss of 
generality. The coefficients $a_{2 n+1}$, $n \geq 1$ are assumed to 
be smaller than $\epsilon^{1/3}$  which 
means that the barrier potential should be only weakly asymmetric. 
Only then, the cubic equation (\ref{kub1}) has real solutions $Q_1$
at most of order $\epsilon^{-2/3}$ for all endpoints in the barrier
region as assumed in deriving (\ref{kub1}) and correction terms
of order $\epsilon^{2/3}$.
For larger asymmetries global features of the potential 
become relevant already in the vicinity of $T_c$.
\begin{figure}[t]
\begin{center}
\leavevmode
\epsfysize=8cm
\epsffile{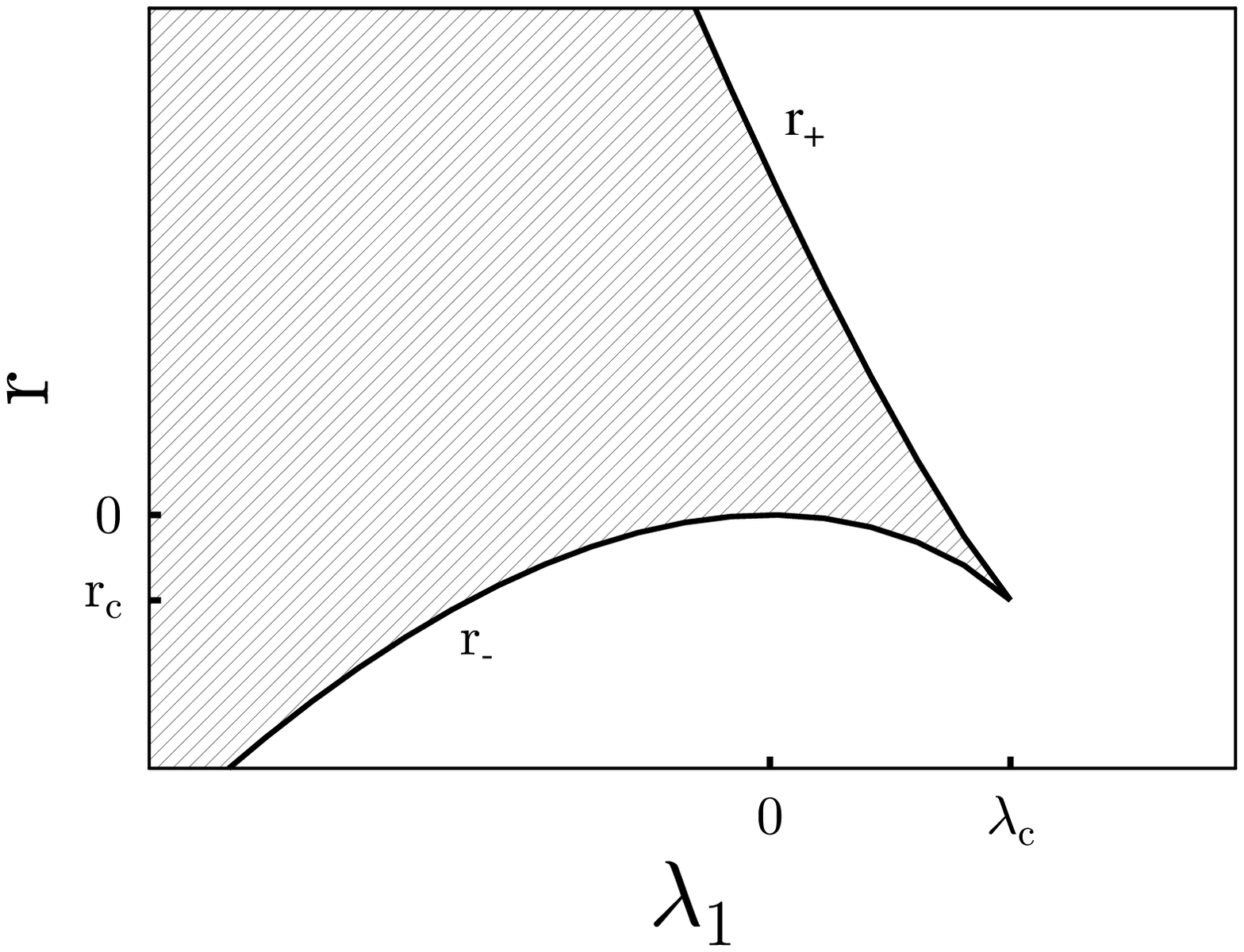}
\end{center}
\caption{The $\lambda_1$-$r$-plane is divided
by the two curves $r_\pm (\lambda_1)$ (solid lines)
into two regions in which the cubic equation (\protect\ref{kub2})
has one or three (shaded region) solutions.}
\label{regio}
\end{figure} 
To make the $\epsilon$-dependence more transparent we set
\begin{equation}
Q=\frac{1}{2 \theta}\epsilon^{2/3} Q_1 .
\label{scale1} 
\end{equation}
\begin{figure}[t]
\begin{center}
\leavevmode
\epsfxsize=13cm
\epsffile{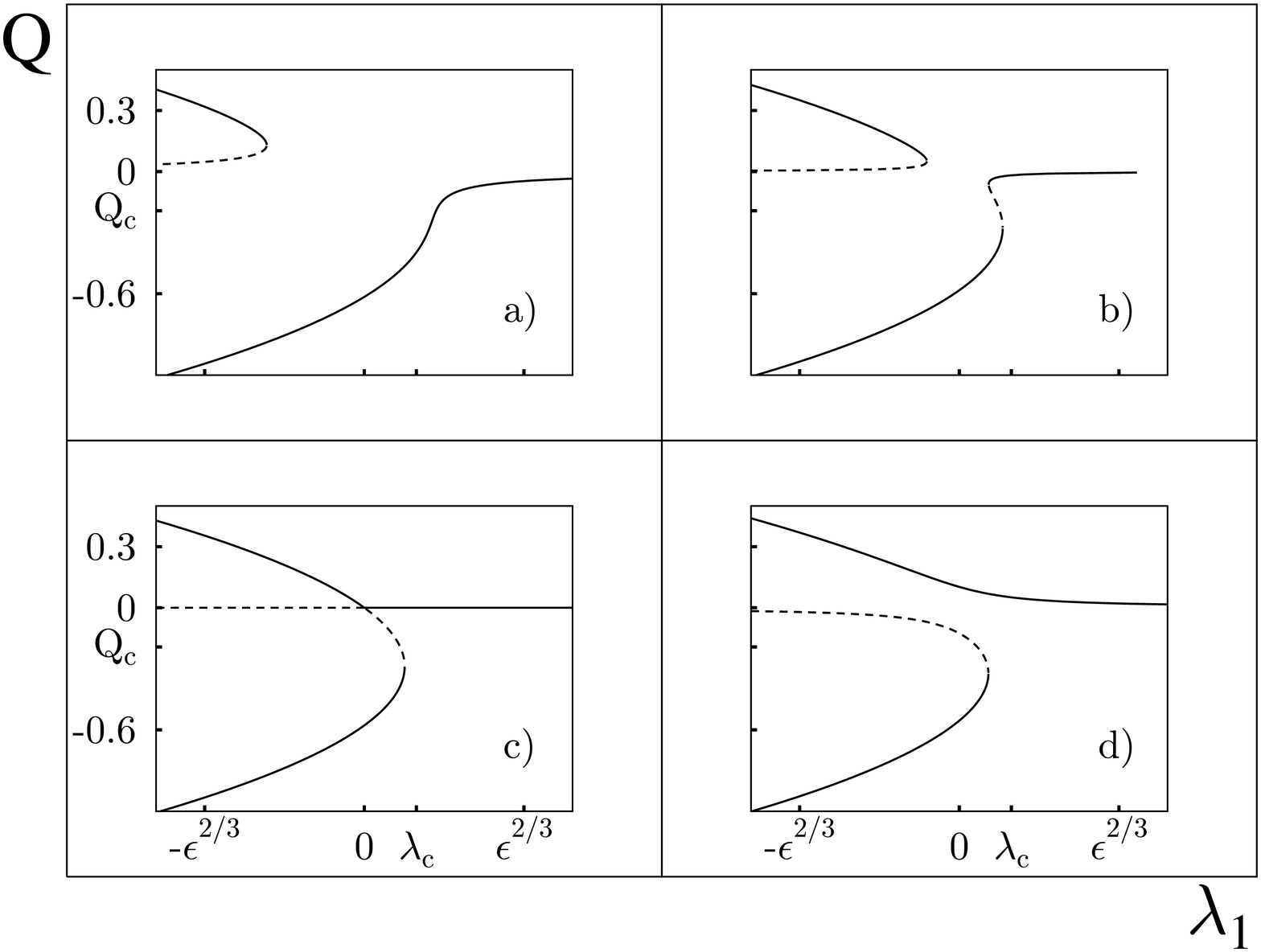}
\end{center}
\caption{Bifurcation scenario for the classical paths for various
boundary values $r$,\protect{ a) $r<r_c$, b) $r_c<r<0$, c) $r=0$, 
d) $r>0$. The solid (dashed) lines represent the stable 
(unstable) solutions of (\protect\ref{kub2}). The potential parameter 
are $a_3 = 1/5$, $a_4=1$ and $\epsilon = 0.01$.}}
\label{bif}
\end{figure}
Then, using  $D_{111}=8/3\pi$, $D_{1111}=3/4$, and $b_1=4 \pi 
r/\theta $, the cubic equation for the mode amplitude $Q$ 
takes the form
\begin{equation}
3 a_4 Q^3 +\frac{16}{3\pi}a_3 \epsilon^{-1/3} Q^2 +\lambda_1
\epsilon^{-2/3}Q =\frac{2\pi}{\theta^2}~r.
\label{kub2}
\end{equation}
From a mathematical point of view (\ref{kub2}) describes a pure 
($r=0$) or a perturbed ($r\neq 0$) one--dimensional bifurcation 
problem with bifurcation parameter $\lambda_1$ \cite{p:iooss}.
The cubic equation (\ref{kub2}) implies two curves 
\begin{equation}
r_\pm(\lambda_1)=r_c\left[ 3\frac{\lambda_1}{\lambda_{c}}-2\pm
2\left(1-\frac{\lambda_1}{\lambda_{c}}\right)^{3/2}\right]
\label{rpm}
\end{equation}
in the $\lambda_1$-$r$-plane where 
bifurcations occur (Fig.~\ref{regio}). Here, the 
critical point $(\lambda_{c},r_c)$ is given by
\begin{eqnarray}
\lambda_{c} &= & 9 a_4 Q_c^2  \epsilon^{2/3}\nonumber\\[10pt]
r_{c} & =& \frac{ 3a_4 \theta^2 }{2\pi}Q_{c}^3 ,
\label{lcrc}
\end{eqnarray}
where
\begin{equation}
Q_{c}  = - \frac{16 a_3}{27 \pi a_4}\epsilon^{-1/3}
\label{qc}
\end{equation}
is the corresponding amplitude. 
Note that $\lambda_c=(4/3)^4 a_3^2/a_4\pi^2$ is independent of $\epsilon$.
Further, $r_c<0$, and $Q_c $ is of order one or smaller for $a_3,a_4>0$
as assumed above.
 
For $\lambda_1 > \lambda_{c} $ there is only 
one solution $Q$ of (\ref{kub2}) which has the same sign as $r$.
As the temperature is lowered, i.e.\ $\lambda_1$ is decreased,
several cases must be distinguished.
First, for $r < r_c$ the high temperature solution continuously
extents to the region $\lambda_1 < \lambda_c$ 
and the amplitude $|Q|$ grows (Fig.~\ref{bif}a). 
Two new solutions with positive sign emerge when the temperature 
reaches the point where $r_-(\lambda_1)=r$ (Fig.~\ref{regio}). 
One of the new solutions 
is stable and the other unstable.
The amplitude of the stable (unstable) path increases (decreases)
with decreasing $\lambda_1$. 
Second, for $r_c \leq r < 0$ one first reaches a point where 
$r_+(\lambda_1)=r$ when the temperature is decreased (Fig.~\ref{regio}).
At this temperature two new branches 
with negative sign but larger amplitude $|Q|$ appear (Fig.~\ref{bif}b). 
Again one path is stable and the other is unstable .
With decreasing $\lambda_1$ the amplitude $|Q|$ of the 
unstable path decreases and 
approaches the stable solution which extends from the 
high temperature region. 
These two paths vanish when the temperature reaches the value
where $r_-(\lambda_1)=r$.
For temperatures below this point there exist only one stable solution
until we arrive at the second solution of $r_-(\lambda_1)=r$.
Here a stable and unstable path with positive sign 
emerge where the amplitude of the stable (unstable) path
increases (decreases) as the temperature is lowered further.
For $r=0$ (Fig.~\ref{bif}c) we have a double point bifurcation 
scenario described below.
Finally, for $r>0$ (Fig.~\ref{bif}d) two new paths appear at the 
point where $r_+(\lambda_1)=r$. Again one solution is stable and the 
other one is unstable.

In particular, for $r=0$ the solutions of (\ref{kub2}) are 
(Fig.~\ref{bif}c)
\begin{equation}
Q=\left\{ 
          \begin{array}{l@{\quad \quad } l}
          0  &  \\[5pt]
          Q_0\left(1
           -\sqrt{1
           -\frac{\lambda_1}{\lambda_{0}}}~\right)&  
                 \\[5pt]
            Q_0\left(1+\sqrt{1
           -\frac{\lambda_1}{\lambda_{0}}}~\right)&
                 ,\end{array} \right.     
\label{lsg0}
\end{equation}
where $Q_0$ and $\lambda_{0}$ are given by
\begin{eqnarray}
Q_0&=&\frac{3}{2}Q_c\nonumber\\[10pt]
\lambda_{0}&=& \frac{3}{4} \lambda_{c} 
\label{l0q0}
\end{eqnarray}
with $\lambda_{c}$ and $Q_c$ from (\ref{lcrc}) and (\ref{qc}), 
respectively. One can show, and it will emerge below, 
that the trivial solution is stable for $\lambda_1\geq 0$, 
the second solution is stable for $\lambda_1 \leq \lambda_{0}$, 
while the third solution is stable for $\lambda_1 \leq 0$.

\subsection{Classical Action Near $T_c$}

Having determined the classical path we are able to 
calculate the minimal action.
Since (\ref{kub2}) may easily be solved numerically, the 
action will be given as a function of the amplitude $Q$ of the 
large amplitude  mode.
To get explicit values for the action the numerical 
value for $Q$ must be inserted.
Now, inserting (\ref{ansatz}) with the Fourier amplitudes 
determined from (\ref{kub1}) and (\ref{kub2}) into (\ref{action1}) 
and expanding in powers of
$\epsilon $,
we obtain after some algebra
\begin{eqnarray}
S_{\rm cl}(Q,z,r) =&& 
        \frac{r^2}{\theta} \Lambda+\frac{ z^2}{\theta} \Omega
           -\frac{ \pi r}{\theta}\epsilon^{-2/3} Q 
-\left(\frac{8 a_3\theta}{9 \pi\epsilon^{1/3}}Q^3 +
            \frac{3 a_4\theta }{4} Q^4\right) \epsilon^{-2/3}
\nonumber\\[10pt]
&-&\frac{4 a_3 \theta (2 +3 \Lambda)r }{3(4\pi^2-\theta^2)\epsilon^{1/3}}Q^2
-\frac{12 \pi a_4 r}{2 \theta \lambda_3}Q^3
-\frac{2 a_3^2 \theta \Gamma}{\epsilon^{2/3}}Q^4\nonumber\\[10pt]
&+&\frac{2 a_3 a_4 \theta}{\epsilon^{1/3}\lambda_3}Q^5
-\frac{a_4^2\theta}{2 \lambda_3}Q^6
+\frac{256 a_5 \theta}{75 \pi \epsilon^{1/3}}Q^5
+\frac{5 a_6 \theta}{3}Q^6 +{\cal O}(\epsilon^{2/3}).
\label{act2}
\end{eqnarray}
Here we have introduced the coefficients
\begin{eqnarray}
\Lambda &= &-\frac{\theta}{2} \tan\left(\frac{\theta}{2}\right)+
\frac{2 \pi}{\pi^2 -\theta^2}\nonumber \\[10pt]
\Omega &= & \frac{\theta}{8} \cot\left(\frac{\theta}{2}\right)
\nonumber \\[10pt]
\Gamma &= & \frac{3}{4}-\frac{\pi^2}{4\pi^2-\theta^2}+
\left(\frac{4 \pi^2}{4\pi^2-\theta^2}\right)^2
\frac{1}{\theta}\tan\left(\frac{\theta}{2}\right)-\frac{64 \theta^2}
{9 \pi^2\left(\pi^2 -\theta^2 \right)},
\label{log}
\end{eqnarray}
which remain finite in the limit $\theta\rightarrow\pi$, i.e.
\begin{eqnarray}
\lim_{\theta \to \pi} \Lambda &=&\frac{3}{2}\nonumber\\[10pt]
\lim_{\theta \to \pi} \Omega &=& 0\nonumber\\[10pt]
\lim_{\theta \to \pi} \Gamma &=& \frac{5}{12}-\frac{16}{3\pi^2}.
\label{limits}
\end{eqnarray}
The first three terms in (\ref{act2}) reduce to the usual 
harmonic action for 
temperatures well above $T_c$. Indeed, for $\lambda_1$ 
larger than order $\epsilon^{2/3}$ the anharmonic terms in 
(\ref{kub2}) can be neglected and we have
\begin{equation}
Q = \frac{2\pi r}{\theta^2\lambda_1}\epsilon^{2/3}+
{\cal O}\left(\frac{\epsilon^{2}}{\lambda_1^{3}}\right)
\label{qht}
\end{equation} 
and hence
\begin{equation}
r^2 \Lambda -\pi r\epsilon^{-2/3} Q =
-\frac{\theta}{2} r^2 \tan\left(\frac{\theta}{2}\right)\left[ 1+
{\cal O}\left(\frac{\epsilon^{4/3}}{\lambda_1^{2}}\right)\right].
\label{sht}
\end{equation}
While the harmonic action diverges for $\theta \rightarrow 
\pi $, i.e.\ $\lambda_1 \rightarrow 0 $, the full action 
(\ref{act2}) remains finite
due to the contributions of the anharmonic terms.

In the special case $q=q'=0$, i.e.\ $r=x=0$, one sees that 
for high temperatures the minimal action vanishes. 
With decreasing 
$\lambda_1 $ two new paths emerge at $\lambda_1=\lambda_{0}$.
At the bifurcation point the amplitudes of the new paths coincide,
and to leading order the action is given by
\begin{eqnarray}
 S_{\rm cl}(Q_0,0,0)_{\lambda_1 =\lambda_{0}}=
\frac{\theta}{4}a_4 Q_0^4\epsilon^{-2/3}   
 + {\cal O}\left(1\right).
\label{s10}
\end{eqnarray}
Since $S_{\rm cl}(Q_0,0,0)_{\lambda_1=\lambda_0}>0$, 
the trivial solution 
($Q=0$ branch) is absolutely stable.
At lower temperatures, i.e.\ smaller $\lambda_1 $, the                    
solutions separate. Denoting the stable and unstable branch by $Q_s$
and $Q_u$, respectively, the new amplitudes reach
$Q_{s}= 2 Q_0$ and $Q_{u} = 0$,
at $\lambda_1 =0$, and the corresponding actions are given by
\begin{eqnarray}
S_{\rm cl}\left(Q_{s},0,0\right)_{\lambda_1 =0}&=&- 4\theta a_4 
Q_0^4 \epsilon^{-2/3}  + {\cal O}\left( 1\right) \nonumber \\[5pt]
S_{\rm cl}\left(Q_{u},0,0\right)_{\lambda_1 =0}&=&0.
\label{sl0}
\end{eqnarray}
Now, $S_{\rm cl}(Q_s,0,0)_{\lambda_1=\lambda_0} < 0 $,
and the $Q_s$ branch is absolutely stable.
Between $\lambda_1=\lambda_0$ and $\lambda_1=0$
at the point $ \lambda_1= 2\lambda_{c} /3 $ 
the two stable paths have to leading  order the same action
and they exchange global stability.
 
This example demonstrates the change of stability of 
the classical paths. At high temperatures there is 
only one minimum of the action in function space.
With decreasing temperature a saddlepoint emerges 
at $\lambda_1=\lambda_0$ with an action larger 
than the minimal action. 
This saddlepoint splits into a maximum and a second 
minimum of the action. The original minimum and 
the new minimum exchange global stability at $\lambda_1=2\lambda_{c} /3$.
Finally, the original minimum and the maximum meet in a
double point at $\lambda_1=0$.

For finite $q$ and $q'$ the cubic equation (\ref{kub2}) may easily 
be solved numerically.
Again new extrema of the action functional arise as the temperature 
is lowered.
However, a change of global stability only occurs for small values of $r$.
A more detailed discussion follows in section 4.4.

\section{Quantum Fluctuations}
Based on the classical paths and the associated extremal
actions, we proceed to determine the equilibrium density matrix
by expanding the action about the classical paths 
according to (\ref{eq:p22}).

We split an arbitrary path $q(\sigma)$ into 
the classical path $q_{\rm cl}(\sigma)$
and a fluctuation $y(\sigma)$.
The relevant fluctuations give a contribution to the full action
of order $\hbar$.
Therefore, the semiclassical expansion is only consistent if the 
classical action is also determined at least to order $\hbar$
that is to order 1 in the dimensionless units used above.
In (\ref{act2}) we have neglected contributions to the classical 
action that are smaller than 1 for coordinates of order 1. 
On the basis of this result we may determine the 
semiclassical density matrix in the 
vicinity of the barrier top.
For coordinates that are larger than order 1
a semiclassical expansion is of course still feasible, but mostly one has to 
use numerical methods to go beyond the approximate 
results derived in the previous section. 
The exception are some special potentials, 
for instance the Eckart barrier potential, where the classical 
mechanics can be solved exactly \cite{eckart}.
 
\subsection{ Expansion about the Classical Path}

To evaluate the pathintegral
an arbitrary path is decomposed into
\begin{equation}
q(\sigma)= q_{\rm cl}(\sigma)+ y(\sigma)
\label{path} 
\end{equation}
and one has to expand the action about $q_{\rm cl}$.
If there exist a set ${q_{\rm cl}^\alpha(\sigma)}$ of stable 
classical paths, the action is expanded
about each $q_{\rm cl}^\alpha(\sigma)$.
Then the density matrix (\ref{eq:p19}) may be written as 
\begin{equation}
\rho_\theta(z,r)=\frac{1}{Z}\sum_\alpha 
\exp\left(-S_{cl}[Q_\alpha ,z,r]\right)f(Q_\alpha ,z,r),
\label{den}
\end{equation}
where $S_{\rm cl}[Q,z,r]$ is the scaled 
minimal action (\ref{action})
and $Z$ denotes an appropriate normalization constant.  
$f(Q_\alpha ,z,r)$ is a functional integral 
over the  paths $y(\sigma)$ 
given by 
\begin{equation}
f(Q,x,r)=\int{\cal{D}} [y] \exp \left( -\sum_{m=2}^\infty 
\delta^m S[q_{\rm cl},y]\right),
\label{flukint}
\end{equation}
where $\delta^m S[q_{\rm cl},y]$ are scaled Fr\^{e}chet 
derivatives according to (\ref{eq:p23}).
Since the fluctuations $y(\sigma)$ have to satisfy the
boundary conditions $y(0)=y(\theta) =0$, we use the Fourier representation 
\begin{equation}
y(\sigma)=\frac{1}{\theta }\sum_{k=1}^\infty Y_k \sin (\nu_k \sigma).
\label{y}
\end{equation}
Then, the integration measure becomes
\begin{equation}
\int{\cal{D}} [y] \cdots ~~= \prod_{n=1}^\infty \left( 
\frac{1}{N}\sqrt{\frac{1}{8 \pi \theta}}\int\limits_{-\infty}^{\infty }
{\rm d}Y_n \cdots \right),
\label{mass}
\end{equation}
where $N$ can be determined from the limit of a free particle, i.e.\ 
$\omega \rightarrow 0$.
With the Fourier representations of the classical path (\ref{ansatz})
and the fluctuations (\ref{y}), respectively, one finds 
for the scaled Fr\^{e}chet derivatives
\begin{eqnarray}
\delta^2S[q_{\rm cl},y] &=&
    \frac{1}{8 \theta}
    \sum_{k=1}^\infty \lambda_k Y_k^2 
    +\frac{1}{8 \theta} \sum_{n=3}^\infty 
    a_n (n-1)\left(\frac{\epsilon}{\theta}\right)^{n-2}  \nonumber \\[10pt]
   && \times\sum_{k_1 \dots k_n =1}^\infty 
    D_{k_1 \dots k_n}~ Q_{k_1}\cdots Q_{k_{n-2}}
    Y_{k_{n-1}} Y_{k_n}   
\label{d2s}
\end{eqnarray}
and 
\begin{eqnarray}
\delta^mS[q_{\rm cl},y] &=&   \frac{1}{4 \theta}
    \sum_{n=m}^\infty a_n \frac{(n-1)!}{m!(n-m)!}
     \left(\frac{\epsilon}{\theta}\right)^{n-2}
      \nonumber\\[10pt]
    && \times\sum_{k_1 \dots k_n =1}^\infty 
    D_{k_1 \dots k_n}~ Q_{k_1}\cdots Q_{k_{n-m}}
    Y_{k_{n-m+1}}\cdots Y_{k_n} 
\label{dms}        
\end{eqnarray}
for $ m\geq 3$.
Here, the coefficients $\lambda_k$ and $D_{k_1\dots k_n}$ 
are defined in (\ref{eigenw1}) and (\ref{d's}), respectively. 
The first term of (\ref{d2s}) arises from the bilinear 
terms of the action, while the remaining terms 
are due to the anharmonicity of the potential.
Clearly, from (\ref{dms}) we see that the semiclassical 
expansion proceeds in powers of $\epsilon$.
Of course, the fluctuation integral (\ref{flukint}) cannot be done
exactly. However, for small $\epsilon$ we may evaluate
the functional integral (\ref{flukint}) perturbatively.
To do so, we first have to solve (\ref{bewgl}) perturbatively
for small $\epsilon$. 
In particular, this yields the order of 
magnitude of the Fourier coefficients $Q_k$ that 
depends on the temperature range.
Afterwards, one estimates the size of the Fourier coefficients 
$Y_k$ of the relevant fluctuations.
The exponent of the integrand of (\ref{flukint}), given by (\ref{d2s}) and 
(\ref{dms}), is an expansion in powers of $Y_k$ with 
coefficients depending on the parameter $\epsilon$, 
if we express the order of magnitude of the $Q_k$ in terms of $\epsilon $.
In a first step the size of the $Y_k$ is estimated from the 
assumption that the terms in the action quadratic in $Y_k$
are of order 1.
However, this estimate is only correct if the terms of higher 
order in $Y_k$ vanish in the limit  
$\epsilon \rightarrow 0$.
If there are higher order terms that grow in this limit
one gets a new (smaller) estimate of the $Y_k$ by requiring
that the leading order stabilizing terms are of order 1.
Again the estimate depends on the range of temperatures 
considered.
\subsection{High Temperatures}
Above, when studying the classical mechanics we have seen that for 
temperatures $\theta \ll \pi$ and for coordinates near the barrier 
top the anharmonic terms are neglectible. 
In the limit of small $\epsilon $ and for high temperatures the magnitude
of the Fourier coefficients $Q_k $ is of order $\theta $.
Assuming that the coefficients $Y_k$ are of order $\theta^{3/2}$
or smaller we see that the first term on the right hand side of 
(\ref{d2s}) is of order 1, while the remaining terms of (\ref{d2s})
and (\ref{dms}) vanish in the limit 
$\epsilon \rightarrow 0$.
Therefore, for the relevant fluctuations the exponent of 
(\ref{flukint}) is given by
\begin{equation}
 S[q]-S[q_{cl}]
    =\frac{1}{8 \theta}\sum_{n=1}^\infty 
\lambda_n Y_n^2 +{\cal O}(\epsilon \theta^2).
\end{equation}
Thus, the path integral (\ref{flukint})
reduces to a product of independent Gaussian integrals
\begin{equation}
f=\frac{1}{N}\prod_{k=1}^\infty \left(\sqrt{\frac{1}{8 \pi \theta}}
\int\limits_{-\infty}^\infty {\rm d}Y_k \exp\left( -\frac{1}{8 \theta}
\lambda_k Y_k^2\right)\right),
\label{flukdet}
\end{equation} 
where the eigenvalues $\lambda_k $ of the second variational operator 
are given by (\ref{eigenw1}).
Now, the fluctuation prefactor for the equilibrium 
density matrix is obtained as
\begin{equation}
f=\left[ 4 \pi \sin(\theta)\right]^{-1/2}
\label{prefharm}
\end{equation}
where the normalization in (\ref{flukdet}) is derived from the free
particle limit.
Combining the action (\ref{harmact}) and the fluctuation prefactor
(\ref{prefharm}), the equilibrium density matrix for coordinates
near the barrier top and for high temperatures emerges as
\begin{equation}
\rho_\theta(z,r)=\frac{1}{Z}
\sqrt{\frac{1}{4 \pi \sin(\theta)}}
\exp\left[- \frac{r^2}{2} \tan\left(\frac{\theta}{2}\right)+
\frac{z^2}{8} \cot\left(\frac{\theta}{2}\right)\right].
\label{densharm}
\end{equation}
Hence, in this temperature and coordinate range, we regain
the well known result for the inverted harmonic oscillator.
Clearly  for $\theta \rightarrow \pi $, this result becomes singular.
Then, the  mode $Y_1$ undergoes large fluctuations
and one has to go beyond the simple semiclassical approximation.

\subsection{Quantum Fluctuations Near $T_c$ }
Since the eigenvalue $\lambda_1$ of the second variational operator vanishes
for $T=T_c$, the simple semiclassical 
approximation fails near $T_c$. 
Under the assumptions made in section 3.2 about the potential parameters
$a_n$ we were able to solve the classical equation of motion 
(\ref{bewgl}) near the critical temperature perturbatively
for small $\epsilon$.
In this temperature range all amplitudes $Q_k$ for $k>1$ are
of order 1 or smaller, while the magnitude of the amplitude 
$Q_1$ is of order $\epsilon^{-2/3}$ or smaller.
We now estimate the magnitude of the Fourier coefficients $Y_k$.
The amplitudes $Y_k$ for $k>1$ are of order 1 or smaller
since all eigenvalues $\lambda_k$ for $k>1$ are of order 1
in this temperature range.
Hence, the fluctuations $Y_k$ for $k>1$ are bounded by the
first term on the right hand side of (\ref{d2s}).

Near the critical temperature the amplitude $Y_1$ can become 
much larger than 1.
To estimate the size of $Y_1$ we first consider the terms
of (\ref{d2s}) that are quadratic in $Y_1$. With (\ref{lcrc})
and (\ref{qc}) we have
\begin{eqnarray}
\delta^2S\left[ q_{\rm cl},y\right]
&-&\frac{1}{8 \theta }\sum_{k=2}^\infty
\lambda_k Y_k^2 =  \nonumber \\[10pt]
~~~~~~~~&~& \frac{1}{8\theta }
\left(\lambda_1-\lambda_{c}+
9 a_4 \epsilon^{2/3}\left(Q-Q_c\right)^2 \right)
Y_1^2 \left[1+{\cal O}\left(\epsilon^{1/2}\right)\right].
\label{quady1}
\end{eqnarray}
From (\ref{quady1}) we see that the coefficient 
\begin{equation}
\Lambda_1(Q)= \lambda_1-\lambda_{c}+
9 a_4 \epsilon^{2/3}\left(Q-Q_c\right)^2
\label{L}
\end{equation}
vanishes at the critical point, where $\lambda_1 =\lambda_c$
and $Q=Q_c$.
Thus the size of $Y_1$ is only bounded by  
higher order terms in $Y_1$.
If we assume that $Y_1$ is of order $\epsilon^{-1/2}$ or smaller
near the critical point, the fourth order variational term
gives
\begin{equation}
\delta^4S\left[q_{\rm cl},y\right]=
\frac{3 a_4}{64 \theta^3 }\epsilon^2 Y_1^4 +o(1).
\label{quarty1}
\end{equation}
Indeed, this term limits the size of 
the mode amplitude $Y_1$ near 
$\lambda_1 =\lambda_c $ to the assumed order of magnitude. 
With this estimate for $Y_1$
and in a narrow range around the critical point,
where $\Lambda_1(Q)\leq {\cal O}(\epsilon)$ one gets for 
the action of the fluctuations  
\begin{equation}
  S[q]-S[q_{cl}]
    =\frac{1}{8 \theta}\sum_{n=2}^\infty 
\lambda_n Y_n^2
+ V(Q,Y_1) +{\cal O}(\epsilon^{1/2}),
\end{equation}
with the fluctuation potential
\begin{eqnarray}
V(Q,Y_1)= \frac{1}{8 \theta}\Bigg[\left(\lambda_1-\lambda_{c} +
9 a_4 \epsilon^{2/3}\left(Q-Q_c\right)^2\right)Y_1^2 \hskip1.5cm  
\nonumber\\[10pt]
 \hskip1.5cm  +3 a_4 \frac{\epsilon^{4/3}}{\theta}
\left(Q-Q_c\right)Y_1^3+
\frac{3 a_4 }{8}\left(\frac{\epsilon}{\theta}\right)^2 Y_1^4 \Bigg].
\label{flukpot}
\end{eqnarray}
From (\ref{L}) we see that the coefficient $\Lambda_1(Q)$
of the harmonic term may vanish. This is 
indeed the case for values of $r$ and $\lambda_1$
on the two curves $r^\pm (\lambda_1)$ introduced in (\ref{rpm}). 
When these curves are crossed, the classical equation of 
motion (\ref{kub2}) has a bifurcation.
Near the bifurcation the remaining terms of the fluctuation 
potential always constrain $Y_1$ to fluctuation amplitudes 
of order $\epsilon^{-1/2}$ as assumed above.
For $\Lambda_1(Q)>0$ the fluctuation potential has only one 
minimum at $Y_1=0$, corresponding to one solution of the 
classical equation of motion.
For $\Lambda_1(Q) \leq 0$ the fluctuation potential has 
three extrema, at $Y_1=0$ and at
\begin{eqnarray}
Y_\pm = \frac{-3 \theta }{ \epsilon^{2/3}} 
\left( Q-Q_c \right) \pm
\sqrt{-\frac{4 \theta^2}{3 a_4 \epsilon^2} 
\left(\lambda_1-\lambda_{c}\right)
 - \frac{ 3 \theta^2}{\epsilon^{4/3}}
\left(Q-Q_c\right)^2 }~.
\label{extrema}
\end{eqnarray}
\begin{figure}[t]
\begin{center}
\leavevmode
\epsfysize=8cm
\epsffile{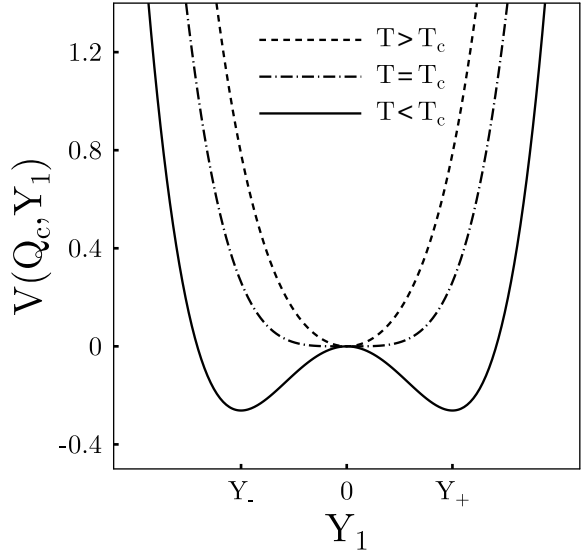}
\end{center}
\caption{The fluctuation potential (\protect\ref{flukpot}) for 
\protect{$Q=Q_c$} and  
temperatures near \protect{$T_c$}.}
\label{flukplot}
\end{figure} 

Specifically, let us discuss the fluctuation potential for
fixed $Q=Q_c$.
Hence, according to (\ref{kub2}), we consider values 
along a line  $r=r_e(\lambda_1)$ in the
$\lambda_1$-$r$-plane, where
\begin{equation}
r_e(\lambda_1)=r_c\left(3\frac{\lambda_1}{\lambda_{c}}-2\right).
\label{line}
\end{equation}
For fixed $Q=Q_c$ the fluctuation potential has the simple form
\begin{equation}
V(Q_c,Y_1)=\frac{1}{8 \theta }\left[\left(\lambda_1-\lambda_{c}
\right)Y_1^2
+\frac{3}{8}a_4 \left(\frac{\epsilon}{\theta}\right)^2Y_1^4\right],
\label{symmpot}
\end{equation}
which is symmetric about $Y_1=0$ (Fig.\ \ref{flukplot}).
According to (\ref{extrema}) for $\lambda_1<\lambda_{c}$  
the extrema $Y_\pm $ are at
\begin{equation}
Y_\pm =\pm\sqrt{\frac{4 \theta^2}{3 a_4 \epsilon^2}(\lambda_{c}-
\lambda_1)}.
\end{equation}
For $\lambda_{1}-\lambda_{c}> \epsilon $ the fluctuation potential 
has only one minimum $Y_1=0$ and the second order variational 
operator constrains $Y_1$ to fluctuation amplitudes smaller than 
order $\epsilon^{-1/2}$.
In this region the simple semiclassical approximation, i.e.\ the 
truncation after the quadratic term in the fluctuation potential, 
is valid.
For $\lambda_1$ in the region 
$0< \lambda_1-\lambda_{c} <\epsilon $,
the curvature of the fluctuation potential at the minimum 
at $Y_1=0$ becomes smaller and 
the fluctuation amplitude of the marginal mode is stabilized 
only by the quartic term. 
When $\lambda_1 $ is decreased further, the minimum at $Y_1=0$ becomes 
a maximum and new minima arise at $Y_\pm$  according to the 
bifurcation scenario discussed for the classical paths.
Indeed, one can show that the extrema of the fluctuation potential 
$V(Q,Y_1)$ are the solutions of the cubic equation  
(\ref{kub2}) for $Q$ apart from the scaling 
factor $2 \theta \epsilon^{-2/3}$ introduced in (\ref{scale1}).
In the region  $-\epsilon < \lambda_1-\lambda_{c}< 0 $ 
the new minima are not well separated by the 
local maximum at $Y_1=0$ and fluctuations 
from one minimum to the other occur.
Hence, in this region the fluctuations are also stabilized by 
the quartic term in the fluctuation potential.
When the temperature is lowered further the barrier height of the local 
maximum of the fluctuation potential becomes larger and 
the fluctuations about the minima decrease.
Then the system is stabilized near the stable 
classical paths and a Gaussian approximation for 
the integral over the fluctuations around these 
paths is again appropriate.

\subsection{The Density Matrix Near $T_c$}
Well above $T_c$ the density matrix is given by (\ref{densharm}).
As $T_c$ is approached the fluctuations grow, and one has to 
take into account the full fluctuation potential (\ref{flukpot}) 
which stabilizes the fluctuations by the quartic term.
In the vicinity of the critical point $(\lambda_c,r_c)$ 
the various solutions of the cubic equation (\ref{kub2}) 
are not well separated and the question arise, which 
branch must be inserted into the action and the 
fluctuation potential.
However, one can easily show that the path integral over the fluctuations 
is independent of the branch chosen.
Expanding the action around the classical paths one is left with the 
fluctuation path integral (\ref{flukint}). 
Using (\ref{mass}) and (\ref{flukpot}) it may be written as
\begin{eqnarray}
f(\theta)&=&\prod_{n=1}^\infty \left( 
\frac{1}{N}\sqrt{\frac{1}{8 \pi \theta}}\int\limits_{-\infty}^{\infty }
{\rm d}Y_n  \right)\exp\left(-\left[\frac{1}{8\theta}\sum_{n=2}^\infty
\lambda_n Y_n^2 + V(Q,Y_1)\right]\right)\nonumber\\[10pt]
&=& \sqrt{\frac{\lambda_1}{4 \pi \sin(\theta)}} K(Q),
\label{flukint2}
\end{eqnarray}
where
\begin{equation}
K(Q)=\sqrt{\frac{1}{8 \pi \theta}}\int\limits_{-\infty}^\infty {\rm d}Y_1
\exp\left( -V(Q,Y_1)\right)
\label{k}
\end{equation}
gives the contribution of the marginal mode $Y_1$.
Now, with (\ref{act2}) and (\ref{k}) the equilibrium density matrix in the 
vicinity of the critical point $(\lambda_{1c},r_c)$ reads
\begin{equation}
\rho_{\theta}(z,r)=\frac{1}{Z}\sqrt{\frac{\lambda_1}{4 \pi \sin(\theta)}}K(Q)
\exp\left(-S_{cl}(Q,z,r)\right).
\label{rho2}
\end{equation} 
Usually, explicit results for the density matrix can be calculated 
only numerically.
After solving the cubic equation of motion (\ref{kub2}) for given 
$\lambda_1$ and $r$ one has to insert $Q$ into the action (\ref{act2}) 
and the fluctuation potential (\ref{flukpot}) and determine 
the fluctuation integral (\ref{k}).
All steps involve rather simple numerics only and some results will 
be presented in section 5.
\begin{figure}[t]
\begin{center}
\leavevmode
\epsfysize=8cm
\epsffile{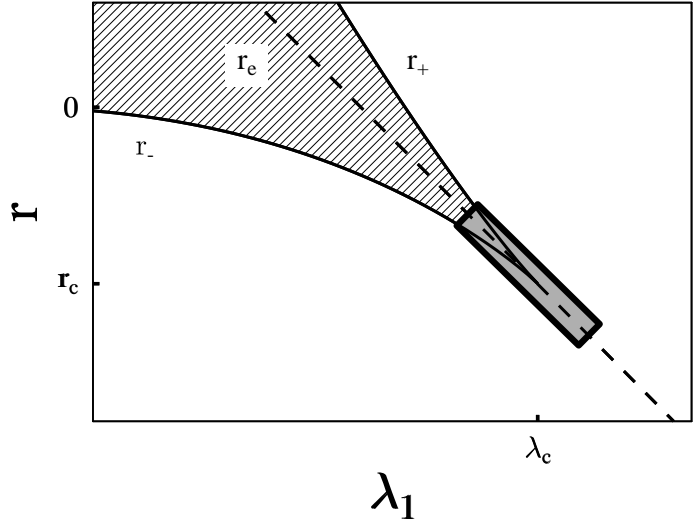}
\end{center}
\caption{Regions in \protect{$\lambda_1$-$r$-plane} in which various
results for the density matrix are valid. Around the critical point 
\protect{$(\lambda_c,r_c)$ in the 
dark region result (65) has to be used. 
In the white region, where only one solution of (30) exists, 
result (67) is appropriate, while in the shaded region result
(70) holds. Along the dashed line $r_e(\lambda_1)$ introduced in
(60) one has $Q=Q_c$} }
\label{r3}
\end{figure} 
As we have seen in section 4.3 the full fluctuation potential 
is needed when the coefficient of the second order term 
$\Lambda_1(Q)$ is of order $\epsilon $ or smaller.
For $|\Lambda_1(Q)|$ larger than order $\epsilon $ the Gaussian 
approximation for the fluctuation potential is appropriate.
Moving away from the critical point $(\lambda_{c},r_c)$, where 
$\Lambda_1(Q)=0$, we reach values of  $|\Lambda_1(Q)|$ of order 
$\epsilon $ by varying $\lambda_1$ by $\Delta\lambda_1$ of order 
$\epsilon $ or by varying $Q$ by $\Delta Q$ of order 
$\epsilon^{1/6}$.
The corresponding  variation $\Delta r$ of the coordinate $r$ 
is then determined  by the cubic equation (\ref{kub2}). 
In the first case, varying $\lambda_1$ for fixed $Q=Q_c$, 
we get $\Delta r ={\cal O}\left(a_3 \right)$.
In the other case, varying $Q$ for fixed $\lambda_1=\lambda_{c}$, 
the order of magnitude of the variation of $r$ is 
$\Delta r ={\cal O}\left(\epsilon^{1/2}\right)$.
Since $Q=Q_c$ along the line $r_e(\lambda_1)$ introduced in
(\ref{line}), the region around the critical point where the 
usual WKB approximation breaks down is of the form of the 
dark rectangle shown in Fig.~4. 
In this region the result (\ref{rho2}) must be used. 

For $\Lambda_1(Q)>\epsilon $ the fluctuation potential can be 
simplified
to read 
\begin{equation}
V(Q,Y_1)=\frac{1}{8 \theta}\Lambda_1(Q)Y_1^2 +o(1),
\label{flpot2}
\end{equation}
where the terms omitted are smaller than order 1 since the 
fluctuations $Y_1$
are now smaller than $\epsilon^{-1/2}$.
$K(Q)$ is therefore a Gaussian integral and the density
matrix (\ref{rho2}) reduces to
\begin{equation}
\rho_{\theta}(z,r)=
\frac{1}{Z}\sqrt{\frac{\lambda_1}{4 \pi \sin(\theta) \Lambda_1(Q)}}
\exp\left(-S_{cl}(Q,z,r)\right).
\label{rho1}
\end{equation}
For high temperatures, i.e.\ $\theta\ll\pi$, one sees from
(\ref{sht}) and (\ref{L}) that the result (\ref{rho1}) matches into
the density matrix in the high temperature limit (\ref{densharm}).

On the other hand, as discussed in section 3.1, 
for temperatures below the critical region
the cubic equation has two stable solutions which we call 
here $Q_{s1}$ and $Q_{s2}$ where $Q_{s1}\leq Q_{s2}$.
As we have seen above, for $ \Lambda_1(Q)< - \epsilon $ these two 
branches are well separated in function space, and a Gaussian 
approximation for the fluctuations around these stable paths is 
appropriate. 
Choosing $Q=Q_{s1}$ the fluctuation potential has a minimum 
at $Y_1=0$ and it can be expanded to give
\begin{equation}
V(Q_{s1},Y_1)= \frac{1}{8 \theta}\Lambda_1(Q_{s1})Y_1^2 +o(1),
\label{flpot3}  
\end{equation}
while near the other minimum at $Y_1=Y_+-Y_-$ the fluctuation 
potential takes the form
\begin{eqnarray}
V(Q_{s1},Y_1)& =& S_{cl}(Q_{s2},z,r)-
                     S_{cl}(Q_{s1},z,r)\nonumber \\[10pt]
 & + & \frac{1}{8 \theta}\Lambda_1(Q_{s2})
\left[Y_1-\left(Y_+-Y_-\right)\right]^2 +o(1).
\label{flpot4}
\end{eqnarray}
The density matrix therefore reads
\begin{eqnarray}
\rho_{\theta}(z,r)=
\frac{1}{Z}\sqrt{\frac{\lambda_1}{4 \pi \sin(\theta)}}
\left(
\frac{1}{\sqrt{\Lambda_1(Q_{s1})}}\exp\left(- 
 S_{cl}(Q_{s1},z,r)\right)
\right.\nonumber\\[10pt]
\left. +\frac{1}{\sqrt{\Lambda_1(Q_{s2})}}\exp\left(
- S_{cl}(Q_{s2},z,r)\right)
\right).
\label{rho3}
\end{eqnarray}

This result matches onto (\ref{rho2}) for values of $\lambda_1$
and $r$ where the coefficient $\Lambda_1(Q)$ becomes of order $\epsilon$
or smaller.
When the difference of the actions of the two stable paths 
$|S_{cl}(Q_{s1},z,r)-S_{cl}(Q_{s2},z,r)|$ is of 
order $1$ or larger, only the  path with smaller action 
contributes to the path integral.
The density matrix then reduces to 
\begin{equation}
\rho_{\theta}(z,r)=\frac{1}{Z}\sqrt{\frac{\lambda_1}{4 \pi \sin(\theta)
\Lambda_1(Q_{s})}}
\exp\left(-S_{cl}(Q_{s},z,r)\right),
\label{rho4}
\end{equation} 
where $Q_{s}$ denotes the path with smaller action.

\section{Discussion and Conclusions}

We have studied the equilibrium density matrix of a quantum particle 
near the top of a general anharmonic potential barrier in the 
temperature range where large quantum fluctuations render the 
harmonic approximation of the potential 
insufficient even for coordinates in the vicinity of the barrier top.
We have shown how the density matrix can be calculated by means of 
the path integral technique employing an expansion about classical paths.
In a narrow range around the critical temperature $T_c$ and in the 
vicinity of 
the critical coordinate $r_c$ the Gaussian approximation for the 
quantum fluctuations, i.e.\ the simple WKB expansion, fails and one 
has to take into account higher order variational terms.
The anharmonicity of the potential remains essential  
below the critical temperature.

\begin{figure}[t]
\begin{center}
\leavevmode
\epsfysize=8cm
\epsffile{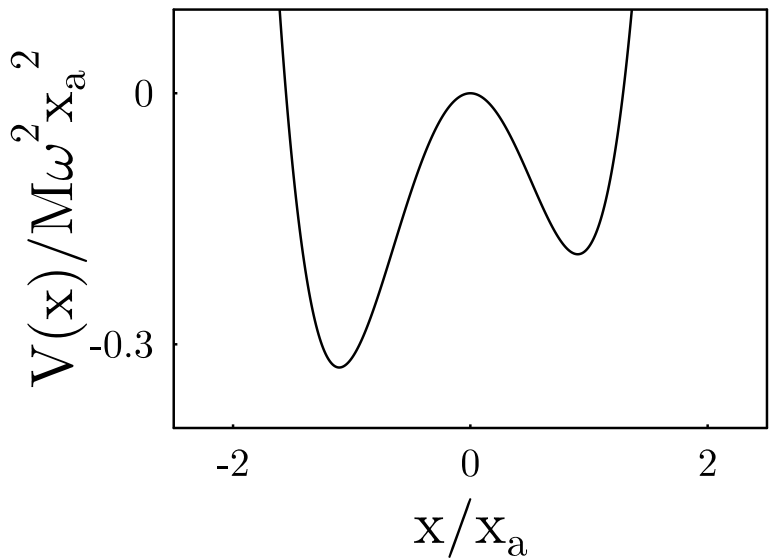}
\end{center}
\caption{The potential (8) with potential parameters
\protect{$ a_3 = 1/5 $}, \protect{$ a_4 = 1 $}, 
and \protect{$a_n=0$} for \protect{$n \geq 5$}.   }
\label{potplot}
\end{figure} 
\begin{figure}[t]
\begin{center}
\leavevmode
\epsfysize=8cm
\epsffile{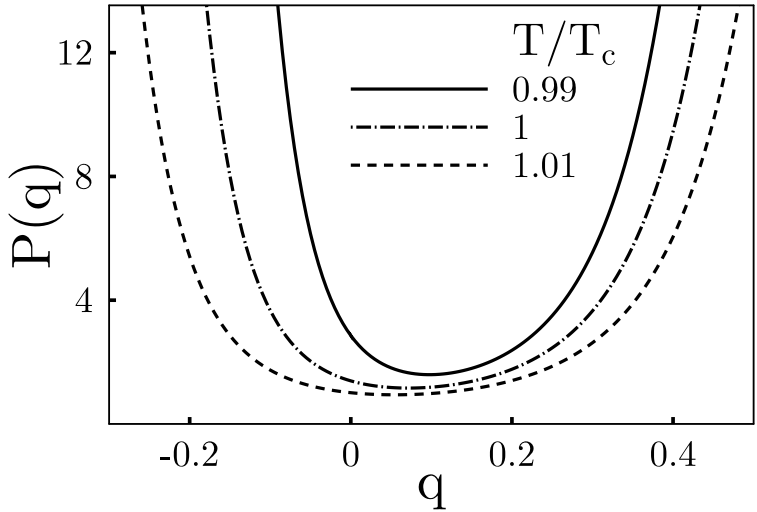}
\end{center}
\caption{The position distribution \protect{$P_\theta (q)$}
 at temperatures near \protect{$T_c$} as a function of 
\protect{$q$} near the barrier top.}
\label{pq}
\end{figure} 
\begin{figure}[t]
\begin{center}
\leavevmode
\epsfxsize=9.5cm
\epsffile{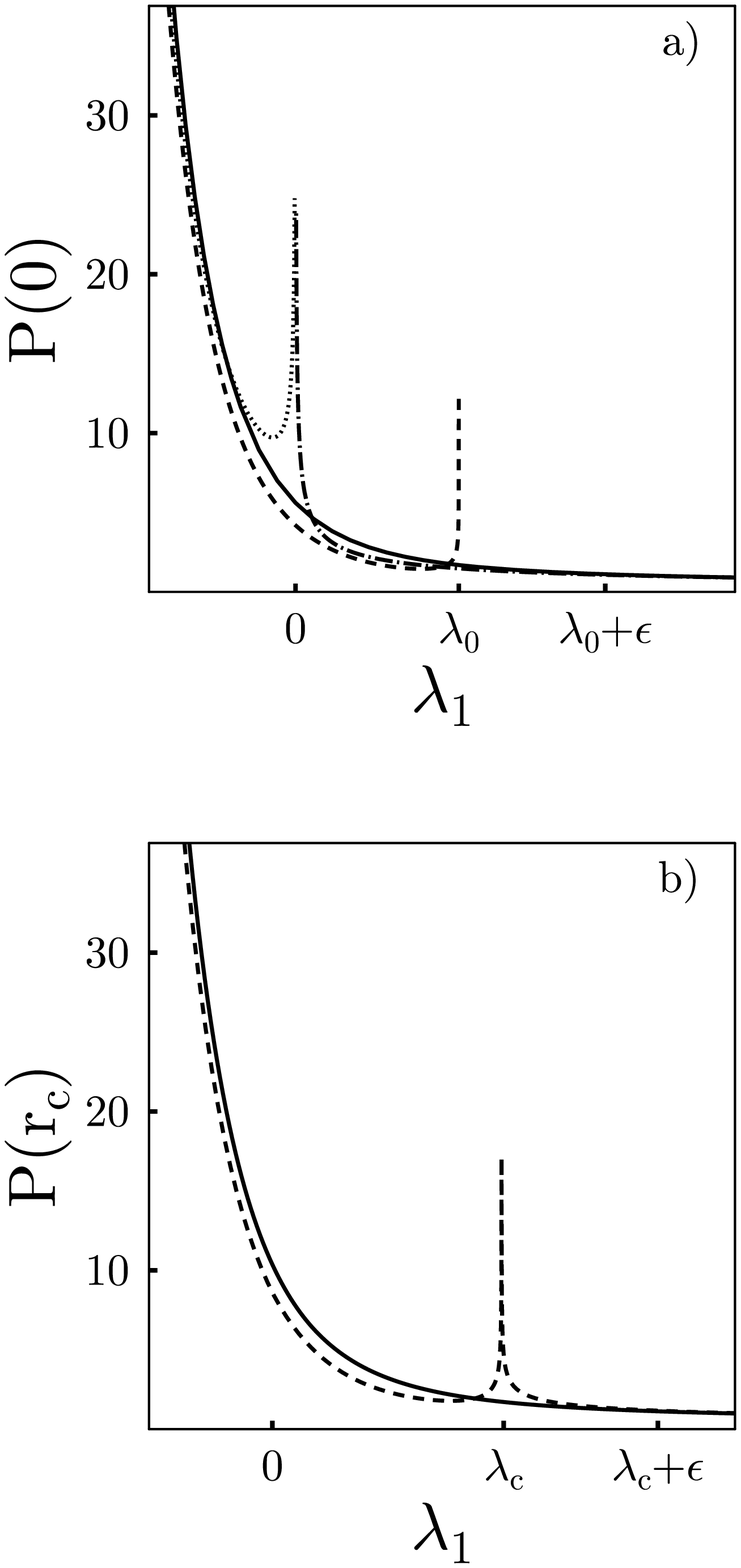}
\end{center}
\caption{The position distribution \protect{$P_\theta (q)$} at 
temperatures near \protect{$T_c$} as a function of the 
eigenvalue \protect{$\lambda_1$} 
for (a)\protect{ $q=0$} and (b) \protect{$q=r_c$}.
The solid line represents the general result (65).
Furthermore, various approximation are shown. 
In (a) the dashed line represents result (71) while the dotted line 
represents result (70). The dotted-dashed line represents
result (67), which is for $q=0$ the same as the high temperature
WKB approximation (53).
In (b) the dashed lines represent result (67).}
\label{plc}
\end{figure}
To illustrate our results we have evaluated the diagonal part of 
the density matrix, i.e.\ the position distribution function  
$P(q)=\rho(z=0,r=q)$, for a system with the potential parameters 
$a_3 = 1/5$, and $a_4 = 1$ while all other coefficients $a_n = 0$. 
Thus the potential (\ref{pot}) reduces to 
\begin{equation}
V(x)=-\frac{1}{2}M\omega^2 x^2\left[ 1-\frac{2}{15}\frac{x}{x_a}
-\frac{1}{2}\left(\frac{x}{x_a}\right)^2\right]
\label{pot1}
\end{equation}
describing an 
asymmetric double well (Fig.~\ref{potplot}).
Using the dimensionless formulation (\ref{coord}) the dimensionless 
potential $\bar{V}=V/\hbar \omega$ reads 
\begin{equation}
\bar{V}(q)=-\frac{1}{4}q^2 + \frac{\epsilon}{30}q^3
+\frac{\epsilon^2}{8} q^4.
\end{equation}
The explicit results  given above 
hold for weakly asymmetric potentials
with $a_3^3 \leq \epsilon \ll 1$. 
This is the case for $ \epsilon = 0.01 $.
  
In Fig.~\ref{pq} the equilibrium position distribution function $P(q)$ is 
depicted for various temperatures near $T_c$.
[The normalization factor for all figures shown in this section 
is choosen as \protect{$Z=1$}.]
For high temperatures the position distribution function is given by 
the harmonic approximation that is symmetric around the
barrier top at $q=0$. 
Anharmonic terms in the equation 
of motion (\ref{kub2}) become relevant with decreasing
temperature.
Accordingly, the minimum of the distribution is shifted
and the distribution becomes more asymmetric 
as the temperature decreases.

In Fig.~7 we show the diagonal part of the density matrix as a 
function of the eigenvalue $\lambda_1$.
At the barrier top, i.e.\ at $q=0$ (Fig.~\ref{plc}a) and for 
high temperatures 
there exist only one solution of (\ref{kub2}),
$Q=0$, and  the distribution function is determined by (\ref{rho1}).
At $\lambda_1=\lambda_0$ a new stable solution $Q_s$
of (\ref{kub2}) emerges.  Between $\lambda_1=\lambda_0$ and $\lambda_1=0$
it exchanges global stability with the $Q=0$ solution and for 
lower temperatures the position distribution function is 
given by (\ref{rho4}).
While the results (\ref{rho1}) and (\ref{rho4}) diverge 
at $\lambda_1=0$ and $\lambda_1=\lambda_0$, respectively, the 
general formular (\ref{rho2}) matches between these results and 
remains finite.
  
For the critical value of the coordinate $q=r_c$ and 
at $\lambda_1=\lambda_c$ there exists only one solution $Q=Q_c$ 
of (\ref{kub2}).
The simple semiclassical approximation (\ref{rho1}) diverges at 
$\lambda_1=\lambda_c$ and we have to take the general result
(\ref{rho2}). In Fig.~7 we see the matching of the result 
(\ref{rho2}) with the result (\ref{rho1}) valid below and 
above the critical region. 

In summary, we have shown how to evaluate the semiclassical density matrix
consistently near the critical region.
For $a_3=0$ one regains the results for the anharmonic 
symmetric barrier potential investigated previously \cite{ankerhold}.
In this case the critical point $(\lambda_c,r_c)=(0,0)$.
A comparison of the semiclassical density matrix with exact results 
will be given elsewhere.

\end{document}